%% file: main.tex
\documentclass[11pt,a4paper]{article}

\usepackage[utf8]{inputenc}
\usepackage[T1]{fontenc}
\usepackage[english]{babel}
\usepackage{lmodern}
\usepackage[margin=2.6cm]{geometry}
\usepackage{graphicx}
\usepackage[export]{adjustbox}
\usepackage{xcolor}
\usepackage{listings}
\usepackage{booktabs}
\usepackage{caption}
\usepackage{enumitem}
\usepackage{microtype}
\usepackage{textcomp}
\usepackage[hidelinks]{hyperref}
\usepackage{url}

\setlist{itemsep=2pt,parsep=0pt,topsep=4pt}
\graphicspath{{figures/}}

\input{listings-mcdc}

\title{\textbf{Integration-First Structural Coverage\\ for Embedded Software:}\\[3pt]
\large Trace-Based Evidence, Hybrid Runtime Analysis,\\ and Cross-Variant Consolidation}

\author{
  Alexander Weiss\thanks{\texttt{aweiss@accemic.com}} \and
  Albert Schulz\thanks{\texttt{aschulz@accemic.com}} \and
  Michael Wittner\thanks{\texttt{Michael.Wittner@razorcat.com}}
}

\date{}

\begin{document}
\maketitle

\begin{center}
\begin{minipage}{0.86\textwidth}
\small
\centering
Alexander Weiss and Albert Schulz are with Accemic Technologies GmbH, Kiefersfelden, Germany.\\
Michael Wittner is with Razorcat Development GmbH, Berlin, Germany.
\end{minipage}
\end{center}

\vspace{4pt}

\begin{center}
\begin{minipage}{0.9\textwidth}
\small\itshape
\textbf{Preprint note.} An earlier version of this work was presented at the embedded
world Conference 2026 (Nuremberg, 11~March 2026) under the title ``Hyper Coverage with
Integration Testing --- How Can I Get Half of My Test Cases for Free?''. The present
document is a revised and restructured author version prepared for open-access archiving;
copyright in the conference version rests with the conference organiser. The technical
content, including all reported measurements, reflects the state of the work as of
March~2026.
\end{minipage}
\end{center}

\vspace{8pt}

\begin{abstract}
\noindent
Structural coverage is widely used as evidence that testing is complete, yet in embedded
projects it is predominantly collected at unit level, simply because that is where
instrumentation and observability are inexpensive. This produces a mismatch. The most
representative completeness signal would come from integration and system tests executed
on the device under test, but classical instrumentation perturbs timing, memory footprint
and concurrency behaviour, while purely trace-reconstructed coverage loses reliability for
decisions and conditions as soon as the compiler optimizes aggressively.

We address this mismatch from both ends. On the process side we describe an
\emph{integration-first} coverage strategy that treats integration and system tests as the
baseline measurement and drives the residual gaps through an explicit closure loop, so
that completeness is established as \emph{covered or justified} rather than as
\emph{covered} alone. On the technical side we use embedded trace as the observation path
and add \emph{hybrid runtime analysis} (hRA): a minimal, semantics-preserving observability
scaffolding that keeps decision and condition boundaries distinguishable in the trace
stream of an optimized (\texttt{-O3}) build, while all coverage state and counting remain
off-target. This converts object-to-source mapping from a heuristic reconstruction into
reviewable evidence and makes branch, condition and MC/DC measurement practical on
release-like binaries. Finally we describe \emph{Hyper Coverage}, a consolidation layer
that merges evidence across test levels, test runs, variants and build configurations, and
that exposes source lines which remain untested in every relevant variant.
\end{abstract}


\vspace{10pt}

\section{Introduction}

A question that recurs in every safety-related or quality-critical embedded project is
when the team may consider itself \emph{done testing}. Functional testing answers one half
of it: whether the requirements appear to be satisfied for the scenarios that were
actually exercised. Structural coverage answers a complementary half: which parts of the
implementation were executed by those tests, and where the blind spots lie.

How much a coverage report is worth is governed by two properties. The first is
\textbf{representativeness} --- whether the binary, the configuration and the execution
environment that were measured correspond to what will ship. The second is \textbf{mapping
integrity} --- whether the correspondence between observed execution and source-level
constructs is repeatable, reviewable, and robust against the transformations a compiler
applies. At unit level both properties are easy to obtain simultaneously. At integration
level they pull in opposite directions: instrumentation tends to preserve mapping
integrity at the cost of representativeness, whereas trace-only measurement preserves
representativeness but degrades mapping integrity precisely for the criteria that matter
most under certification, namely decisions and conditions. Closing that divergence is the
subject of this paper.

In current practice, coverage is frequently treated as a unit-test deliverable. The reason
is pragmatic rather than methodological: unit-test environments permit software
instrumentation and offer high observability at negligible cost. Integration and system
tests, however, execute the representative build in the representative configuration. They
exercise interfaces, concurrency, timing and configuration-dependent behaviour that unit
tests cannot reproduce faithfully, and therefore constitute the more honest readiness
signal.

What prevents projects from acting on this is measurement technology rather than
methodology. Software instrumentation inserts additional instructions and memory traffic,
alters scheduling, and can invalidate real-time behaviour. On multi-core devices it also
shifts contention patterns, which may mask existing concurrency defects or introduce new
ones. Faced with this, many projects fall back to unit-test-centred coverage reporting ---
and integration-level gaps remain masked behind unit-level numbers that look complete.

This paper makes four contributions:

\begin{enumerate}
  \item An \textbf{integration-first coverage process} that adopts integration and system
        tests on the device under test as the baseline completeness metric, and that
        applies a structured gap-closure loop to reach a defensible notion of
        \emph{covered or justified}.
  \item A \textbf{trace-based measurement pipeline} for continuous integration- and
        system-level coverage, which produces object-level execution evidence as the
        primary truth and derives source-level coverage from it through explicit mapping
        artifacts.
  \item \textbf{Hybrid runtime analysis (hRA)}, which preserves decision and condition
        observability under aggressive optimization, turning object-to-source mapping from
        a heuristic into verifiable evidence and thereby enabling robust branch, condition
        and MC/DC measurement on optimized binaries.
  \item \textbf{Hyper Coverage reporting}, which consolidates evidence across test levels,
        test runs, variants and configurations, and which surfaces source lines that
        remain untested in every relevant variant.
\end{enumerate}

The intent is explicitly \emph{not} to abolish unit testing. Unit tests remain
indispensable for fast feedback during development, for refactoring confidence, and for
cases in which integration-level stimulation is infeasible. The goal is narrower and more
economic: to stop writing and maintaining audit-grade unit tests whose only purpose is to
\emph{manufacture} coverage numbers for code that realistic integration tests already
exercise.

\section{Why Coverage at Integration Level Is Hard}

\subsection{What structural coverage establishes --- and what it does not}

Structural coverage records the execution of code elements under a given test suite. The
customary criteria are statement, branch, condition and modified condition/decision
coverage (MC/DC)~\cite{chilenski2001,chilenski1994}. It is worth restating plainly that
coverage is evidence of \emph{execution}, not a proof of \emph{correctness}: a defect that
is executed is a covered defect. Its value lies elsewhere --- as a risk radar that makes
untested regions visible, and as a planning instrument for systematic, gap-driven test
improvement.

A point that is easily lost in tooling discussions is that coverage is meaningful only
with respect to a representative build and configuration. A dedicated coverage build that
deviates from the release build weakens the evidence it produces and complicates the
argument that has to be made towards an auditor.

\subsection{Code that exists but is never built}

Coverage results are conventionally reported per build. In variant-rich product lines this
creates a systematic blind spot: reporting 100\,\% coverage for one configuration says
nothing about code that exists in the source tree but was excluded from that particular
build. Conditional compilation and build switches can hide code that is simply never
active in the measured configuration. From both a safety and a security perspective this
matters, because dormant code may still ship in some other configuration and may contain
defects or vulnerabilities that no measurement ever touched.

\subsection{Test levels and their observation requirements}

Unit tests isolate code and provide controllability. Integration tests validate
interactions, interfaces, timing behaviour and configuration realism. Field experience
suggests that many incidents are not single-function failures at all; they arise from
mismatched assumptions between components. This argues for treating integration coverage
as a first-class completeness metric rather than as an optional supplement to unit
coverage.

The two levels also impose fundamentally different requirements on the observation
mechanism, as summarised in Table~\ref{tab:obs}. At unit level, a measurement may distort
timing and footprint without invalidating its own result. At integration level it may not.

\begin{table}[htbp]
\centering
\caption{Observation requirements by test level.}
\label{tab:obs}
\small
\begin{tabular}{lll}
\toprule
\textbf{Requirement} & \textbf{Unit and module tests} & \textbf{Integration and system tests}\\
\midrule
Long observation windows        & not required & required \\
No change in timing behaviour   & not required & required \\
No change in memory footprint   & not required & required \\
Tolerance towards probe effect  & higher       & very low \\
\bottomrule
\end{tabular}
\end{table}

\subsection{The cost of software instrumentation}
\label{sec:instr-cost}

Software instrumentation typically inserts counters or tags at the points of interest.
Even the elementary \emph{read counter, increment, write back} pattern adds memory traffic
and can introduce non-deterministic bus effects on multi-core devices. For integration
tests that depend on timing realism this is consequential in several distinct ways:

\begin{itemize}
  \item additional instructions change the instruction mix and thereby scheduling;
  \item additional memory operations change cache behaviour and latency;
  \item real-time constraints may be violated, and race conditions may be masked or
        created;
  \item the memory footprint may exceed platform budgets, so that the instrumented build
        cannot be executed on the target at all;
  \item the \emph{coverage build} ceases to be the \emph{release build}, which weakens the
        resulting evidence.
\end{itemize}

In safety-related workflows a dedicated measurement build is sometimes unavoidable in
order to obtain structural evidence at all. Where this is the case, confidence in
representativeness is usually established by validating the measurement build against the
production build --- for instance by executing the same functional test cases side by side
and comparing results across the test lifecycle, as recommended in industrial tool
workflows~\cite{lauterbach2023}.

\begin{figure}[htbp]
\centering
\includegraphics[max width=\textwidth]{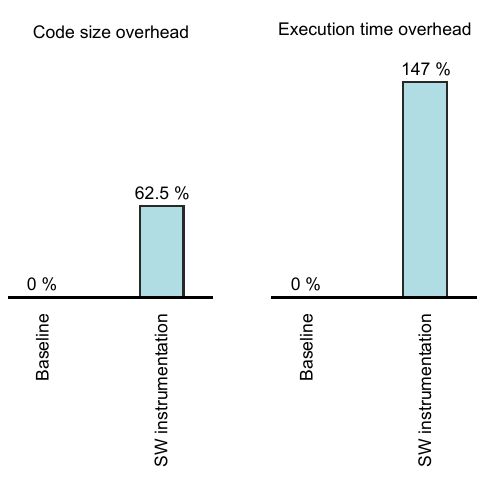}
\caption{Software instrumentation overhead --- a blocker for integration tests. Workload:
SQLite test run; baseline: no instrumentation; instrumented: gcc/gcov; metrics: code size
and execution time delta.}
\label{fig:instr}
\end{figure}

\paragraph{An indicative measurement.}
To illustrate the order of magnitude, Fig.~\ref{fig:instr} contrasts an
instrumentation-based measurement with a trace-based one on a single workload. On an
SQLite test run executed on an Arm Cortex-A53 at 1.5\,GHz (32\,KiB L1-D, 32\,KiB L1-I,
1\,MiB L2), classic gcc/gcov instrumentation~\cite{gcov} increased \textbf{code size by
62.5\,\%} and \textbf{execution time by 147\,\%} relative to an uninstrumented baseline,
whereas the preserved build used for trace-based measurement stayed close to the baseline.

Two qualifications belong with these numbers. First, they are \emph{indicative}: they
describe one workload on one target with one compiler configuration, measured in March
2026, and they are not offered as a general benchmark. Second, and more importantly, the
argument of this paper does \emph{not} rest on the claim that instrumentation is always
expensive. Instrumentation overhead varies strongly with workload, compiler and code
layout; there are workloads in which arc counters are absorbed almost entirely by the
microarchitecture and the measured overhead approaches zero. The argument is a different
one: because the overhead is workload- and layout-dependent, the timing behaviour of the
production binary cannot be bounded from an instrumented build \emph{a priori} --- and
bounding it is precisely what integration-level evidence is supposed to accomplish.

\subsection{Prior foundation}

Earlier work argued that embedded trace can supply non-intrusive monitoring and continuous
coverage at integration level without instrumentation overhead. It also introduced a
structured gap-analysis loop and the notion that completeness may be defined as coverage
by tests \emph{plus} coverage by justification~\cite{dx2024}. The present paper builds on
that foundation and extends it in two directions: towards robust decision and condition
evidence on optimized binaries (Sections~\ref{sec:trace} and~\ref{sec:hra}), and towards
variant-aware consolidation of that evidence (Section~\ref{sec:hyper}).

\section{An Integration-First Coverage Process}

\subsection{Overview}

\begin{figure}[htbp]
\centering
\includegraphics[max width=\textwidth]{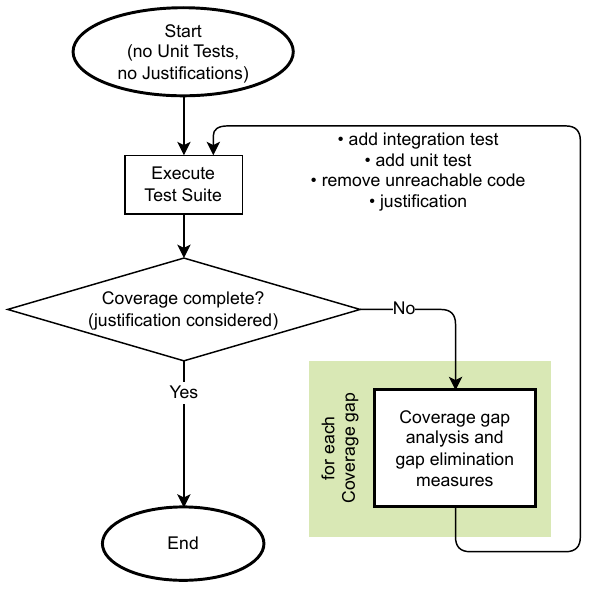}
\caption{The ``integration-first'' coverage process.}
\label{fig:process}
\end{figure}

The process sketched in Fig.~\ref{fig:process} starts from an observation about how
projects are actually organised: requirements-driven integration tests already exist in
most of them. What is missing is not the tests but the conversion of those tests into
structural evidence --- and a discipline that drives the remaining gaps at the \emph{same}
test level, so that integration gaps are not quietly masked by unit tests written to close
them on paper.

\begin{figure}[htbp]
\centering
\includegraphics[max width=\textwidth]{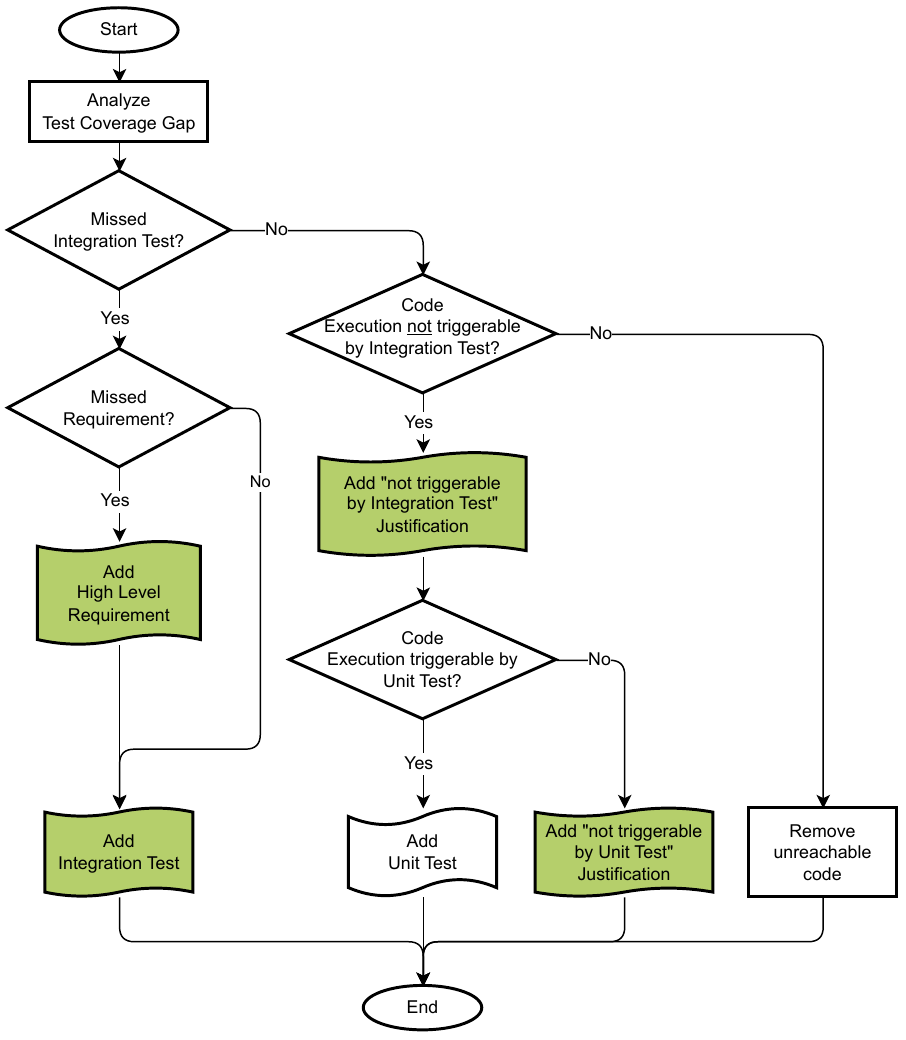}
\caption{Gap analysis process.}
\label{fig:gap}
\end{figure}

The process has three steps:

\begin{enumerate}
  \item \textbf{During development}, unit tests exist for fast feedback, refactoring
        support and module stabilisation. They are useful, but they are not assumed to be
        audit-grade by default.
  \item \textbf{Functional integration and system tests} are executed against requirements,
        on the device under test or in a representative environment, and structural
        coverage is measured continuously while the suite runs.
  \item \textbf{A gap-analysis loop} (Fig.~\ref{fig:gap}) is applied: for every uncovered
        element the team decides whether to add an integration test, to add a missing
        requirement, to remove unreachable code, to justify an infeasible gap, or --- only
        where justified --- to add a complementary unit test.
\end{enumerate}

In practice, well-structured requirements-driven integration and system suites frequently
reach a substantial baseline of structural coverage before any explicit gap-closure
activity begins. The precise figures are strongly project- and domain-specific and are
frequently subject to non-disclosure obligations, so we refrain from stating a general
number. What industrial experience does indicate is a qualitative shift in the role of
unit testing: where integration and system tests already cover a significant fraction of
the implementation, unit tests increasingly serve as \emph{targeted complements} for
scenarios that are infeasible or uneconomic to stimulate at integration level, rather than
as the primary vehicle for manufacturing coverage.

\subsection{Justified gaps and the quality of evidence}

A coverage gap is not automatically a defect in the test suite. Some gaps are justified,
and the justification is itself part of the evidence. Recurring categories include:

\begin{itemize}
  \item defensive programming paths that ought to exist but ought not to be triggered in
        normal operation;
  \item test-environment economics, where extreme conditions are too costly or impractical
        to stimulate;
  \item safety constraints, where a state would be unsafe to provoke on the real system;
  \item variants or features not used in the configuration under assessment;
  \item partially used legacy or third-party library code.
\end{itemize}

\begin{figure}[htbp]
\centering
\includegraphics[max width=\textwidth]{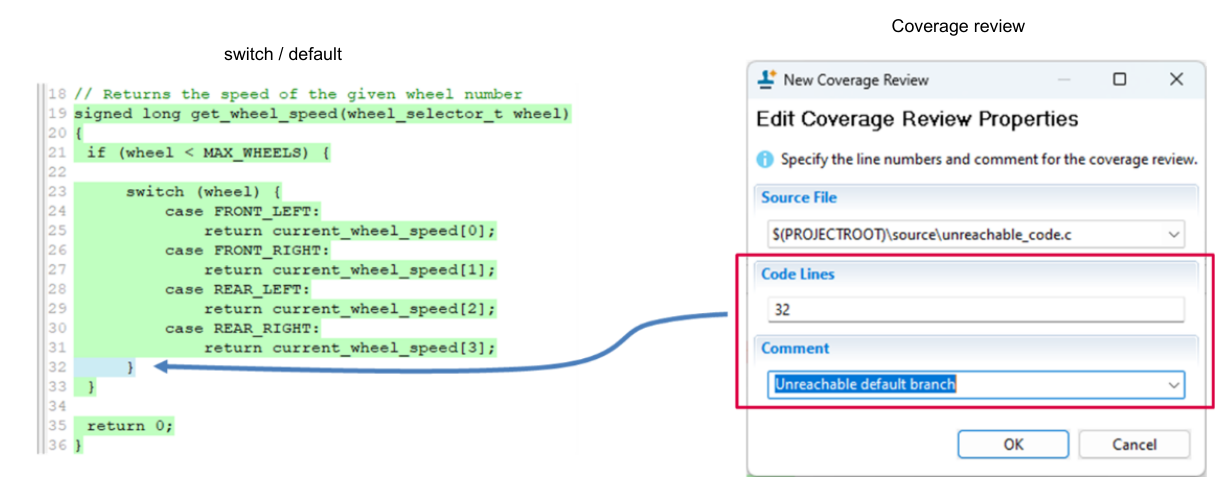}
\caption{Marking ``not reachable code'' and reviewing coverage decisions.}
\label{fig:unreach}
\end{figure}

The governing rule is that every exclusion carries a documented rationale and remains
traceable to the affected code elements (Fig.~\ref{fig:unreach}).

\subsection{Handling unreachable code}

Unreachable code deserves an explicit decision rather than a silent exclusion. If code is
genuinely dead, removing it reduces risk and improves maintainability. If it is
unreachable only under the chosen test level or configuration, the reasoning is documented
and the gap is justified accordingly. Tool support for marking, reviewing and re-reviewing
unreachable code is an important ingredient of audit readiness, because such decisions
must survive later changes to the code base.

\section{Trace-Based Measurement and the Limits of Naive Mapping}
\label{sec:trace}

\subsection{Embedded trace as an observation path}

\begin{figure}[htbp]
\centering
\includegraphics[max width=\textwidth]{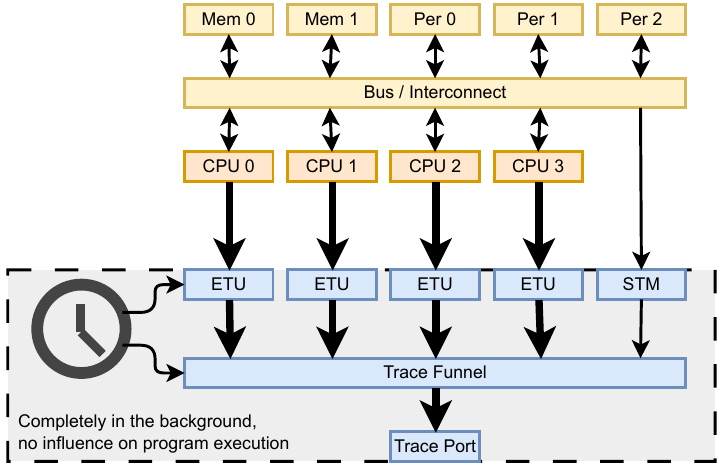}
\caption{Embedded trace overview.}
\label{fig:traceovw}
\end{figure}

Many contemporary processors and SoCs provide hardware trace facilities~\cite{trace2022}
--- for instance Arm\textregistered{} ETM via CoreSight\texttrademark{}~\cite{coresight},
or the trace units of Infineon AURIX\texttrademark{} TriCore\texttrademark{}
microcontrollers (Fig.~\ref{fig:traceovw}). Because trace acquisition is passive, it
permits long observation without modifying the software under test. This makes measurement
on representative builds possible and renders regression comparisons more meaningful,
since no special test-only build has to be produced and maintained.

Two properties of trace shape everything that follows. First, modern trace protocols
principally emit \emph{control-flow changes}: source-level control flow is observable only
where it leaves a corresponding footprint in the object-code control flow. Second, trace
data lives in the object-code domain. Deriving source-level coverage therefore requires a
dependable mapping from executed instruction addresses and control-flow decisions back to
source-level structures.

\subsection{From trace stream to object-level evidence}

A trace coverage pipeline typically proceeds in five stages: capture the trace stream;
reconstruct the executed instruction addresses and control flow; aggregate the evidence
across test runs; produce object-level coverage evidence as the primary truth; and finally
transform the object-level evidence to source level using explicit mapping logic.

If this pipeline can be executed online, the system supports long observation windows and
delivers immediate results --- a prerequisite for system tests that run for hours or days.
Live trace processing, as implemented for example in
CEDARtools\textregistered{}.Coverage~\cite{cedartools}, performs continuous on-the-fly
reconstruction and coverage analysis of the trace stream. This suits integration testing
particularly well, because the observation window is then limited neither by a finite
trace buffer nor by the latency of post-processing very large trace recordings.

\subsection{Where debug-information mapping breaks}

The naive approach maps executed addresses back to source lines and branches using debug
information~\cite{dwarf5}. This works acceptably for statement and line coverage under
modest compiler transformations. It degrades sharply for the more demanding criteria on
optimized builds, for a structural reason: optimizing compilers rewrite exactly the
control flow that the mapping depends upon. They remove and transform branches, invert
conditions, merge blocks, inline functions and move code. The consequences are:

\begin{itemize}
  \item if-conversion, predication and block merging rewrite control flow;
  \item a branch may disappear entirely, so the corresponding decision is no longer
        directly observable;
  \item the direction of a conditional branch (true/false) is chosen by the compiler and
        may be inverted with respect to the source;
  \item inlining and code motion blur which source-level decision caused which observed
        behaviour;
  \item automated identification of decisions and conditions therefore becomes unreliable;
  \item and trace-only mapping may consequently become non-verifiable under optimization,
        which is to say: not auditable as evidence.
\end{itemize}

\paragraph{Example: if-conversion.}
A common case is a decision that exists in the source but not as a branch in machine code.
To improve performance the compiler replaces control flow with predicated or conditional
execution (Fig.~\ref{fig:ifconv}). Trace cannot observe a branch that no longer exists, so
branch-based reconstruction fails --- not because the measurement is imprecise, but
because the event it looks for is absent.

\begin{figure}[htbp]
\centering
\begin{minipage}{0.88\textwidth}
\textit{Source:}
\begin{lstlisting}[style=mcdcasm]
if (a == 0)
    b = 42;
\end{lstlisting}
\smallskip
\textit{Traditionally this if-branch yields a branch instruction, e.g.\ \texttt{bne} on Arm:}
\begin{lstlisting}[style=mcdcasm]
cmp   r3, #0                 # if (a == 0)
bne   10494 <foo()+0x40>     # --- skip mov/str if (a != 0)   (*\annotgreen{add branch to trace}*)
mov   r3, #42 ; 0x2a         # b = 42;
str   r3, [fp, #-12]         # ...
\end{lstlisting}
\smallskip
\textit{But Arm gcc compiles it down to:}
\begin{lstlisting}[style=mcdcasm]
cmp     r3, #0
moveq   r3, #42 ; 0x2a       (*\annotred{not visible in trace}*)
streq   r3, [sp, #4]
\end{lstlisting}
\end{minipage}
\caption{Example of if-conversion to a conditional move (\texttt{moveq}). The decision
still exists in the source, but after if-conversion it is no longer a branch in the
object code, so there is nothing left for the trace to observe.}
\label{fig:ifconv}
\end{figure}

\paragraph{Example: MC/DC under high optimization.}
At high optimization levels, complex Boolean decisions may be lowered into arithmetic, or
into patterns in which the outcomes of the individual conditions are never represented as
explicit branches (Fig.~\ref{fig:mcdcfail}). Trace shows what executed at machine level,
but not the outcome of each atomic condition. MC/DC therefore cannot be reconstructed
reliably from control flow alone.

\begin{figure}[htbp]
\centering
\begin{lstlisting}[style=mcdcasm]
int T2_original (int A, int B, int C) {
    if ((A && B) || C) {
        return 1;
    }
    else {
        return 0;
    }
}

// resulting commented assembly code
//
// 50e  snez  a0, a0      # a0 := (a0 != 0) ? 1 : 0
// 512  snez  a1, a1      # a1 := (a1 != 0) ? 1 : 0
// 516  and   a0, a0, a1  # a0 := a0 & a1 -> logical AND of (a0 != 0) && (a1 != 0)
// 518  snez  a1, a2      # a1 := (a2 != 0) ? 1 : 0
// 51c  or    a0, a0, a1  # a0 := a0 | a1 -> logical OR of previous result with (a2 != 0)
// 51e  ret               # return a0
\end{lstlisting}
\caption{MC/DC under \texttt{-O3}. Trace-only measurement can fail because of the
arithmetic evaluation of the multi-condition, which is not visible to embedded trace as no
control-flow change occurs. The illustrating RISC-V assembly is generated by gcc with
\texttt{-O3}.}
\label{fig:mcdcfail}
\end{figure}

\subsection{The resulting trade-off}
\label{sec:tradeoff}

The state of the art thus presents a trade-off rather than a solution:

\begin{itemize}
  \item software instrumentation provides strong mapping and clear decision semantics, but
        can compromise the representativeness of timing, resource usage and concurrency
        behaviour at integration level;
  \item trace-only measurement preserves the representativeness of the executed binary and
        its environment, but its mapping becomes fragile under optimization.
\end{itemize}

For certification and audit purposes both halves are needed at once: release realism
\emph{and} reliable source-level evidence, particularly for decisions and conditions.

In current industrial and academic practice~\cite{rapicover,lauterbach2023,vector2022},
object-code-based coverage solutions either rely on extensive software instrumentation,
which tends to conflict with integration-level timing and resource constraints, or on
trace-only reconstruction, which works well for line and function coverage but becomes
fragile for decision, condition and MC/DC coverage under aggressive optimization and
modern compiler transformations. The approach presented here is designed to
\emph{complement} these solutions by offering a trace-centric, integration-friendly path
to reliable decision and condition evidence on optimized binaries, and it is intended to
integrate with existing toolchains rather than to replace them.

\section{Hybrid Runtime Analysis}
\label{sec:hra}

\subsection{Design goals}

Hybrid runtime analysis (hRA) is intended to bridge the gap between execution realism and
mapping reliability identified in Section~\ref{sec:tradeoff}. Its design goals are:

\begin{itemize}
  \item retain a release-like integration build, that is, optimized and with realistic
        timing;
  \item preserve source-level control-flow intent at decision and condition boundaries;
  \item produce trace evidence that contains stable anchors for decisions and conditions;
  \item externalise coverage counting and state management off-target, so as to avoid
        cache pollution and counter traffic on the device;
  \item make the mapping verifiable rather than heuristic.
\end{itemize}

The core hybrid runtime analysis concept described in this section is the subject of a
pending international patent application by Accemic Technologies. The intention is to
enable controlled industrial adoption of the technology through integration into
established coverage and test toolchains.

\subsection{Concept: trace observation plus source-code supplementation}

The term \emph{hybrid} refers to the combination of two mechanisms. Trace remains the
observation mechanism; in addition, the build is supplemented with minimal, local
modifications that preserve the observability of decisions and conditions. It is essential
that these modifications are not classical instrumentation. Specifically, they

\begin{itemize}
  \item do not maintain per-point counters in target memory;
  \item avoid frequent read-modify-write patterns; and
  \item are confined to ensuring that decision and condition evaluation events remain
        distinguishable within the trace stream.
\end{itemize}

\begin{figure}[htbp]
\centering
\includegraphics[max width=\textwidth]{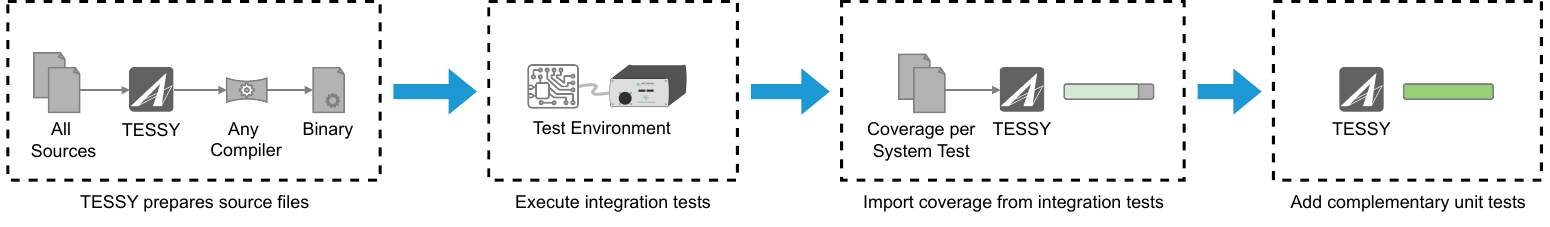}
\caption{Hybrid runtime analysis flow.}
\label{fig:hraflow}
\end{figure}

\subsection{Placement in a test toolchain}

In a practical workflow (Fig.~\ref{fig:hraflow}) the test environment prepares the source;
the integration suite is executed on the device under test while trace is captured,
processed and mapped back to source with CEDARtools\textregistered{}.Coverage; the
resulting coverage evidence is imported into the reporting system; and only afterwards are
complementary unit tests added for those gaps that turn out to be infeasible at
integration level. The ordering matters: unit tests are written \emph{in response to}
identified gaps, not in anticipation of a coverage target.

\subsection{Making MC/DC observable under optimization}

To render MC/DC provable under aggressive optimization, hRA introduces observability
points for each atomic condition of a decision. These points are locally protected against
those specific optimizations that would otherwise erase the decision structure. The trace
then contains distinct, traceable evaluation events for each atomic condition, and the
off-target analysis reconstructs the condition histories that MC/DC requires
(Fig.~\ref{fig:mcdchra}).

The concrete control-flow preservation scheme and the condition observability patterns are
beyond the scope of this paper. The idea that matters here is that a minimal,
semantics-preserving scaffolding survives optimization and turns each relevant condition
evaluation into a stable, trace-visible anchor.

\begin{figure}[htbp]
\centering
\begin{lstlisting}[style=mcdcasm]
int T2_original (int A, int B, int C) {
    if (*\hlanchor{((A \&\& B) || C)}*) {
        return 1;
    }
    else {
        return 0;
    }
}

// resulting commented assembly code
//
// 50e  (*\hlanchor{beqz a0,516}*)  # If A == 0 -> skip to check C
// 510  (*\hlanchor{beqz a1,516}*)  # If B == 0 -> skip to check C
// 512  li a0,1      # return 1
// 514  ret
// 516  (*\hlanchor{beqz a2,51c}*)  # If C == 0 -> skip result = 1
// 518  li a0,1      # return 1
// 51a  ret
// 51c  li a0,0      # return 0 (if C was false)
// 51e  ret
\end{lstlisting}
\caption{MC/DC under \texttt{-O3} with hRA. The control flow for each atomic condition is
preserved and remains observable by means of embedded trace. The illustrating RISC-V
assembly is generated by gcc with \texttt{-O3} after the described control-flow
preservation measures.}
\label{fig:mcdchra}
\end{figure}

\subsection{Why this is not classical instrumentation}

Classical instrumentation collects runtime state by writing to counters or buffers on the
device under test. hRA instead concentrates on \emph{preserving observability} and
performs the counting off-chip. This reduces runtime disturbance and avoids thrashing of
counter arrays, while retaining transparency, because the modifications remain explicit at
source level and can be reviewed there.

\begin{figure}[htbp]
\centering
\includegraphics[max width=\textwidth]{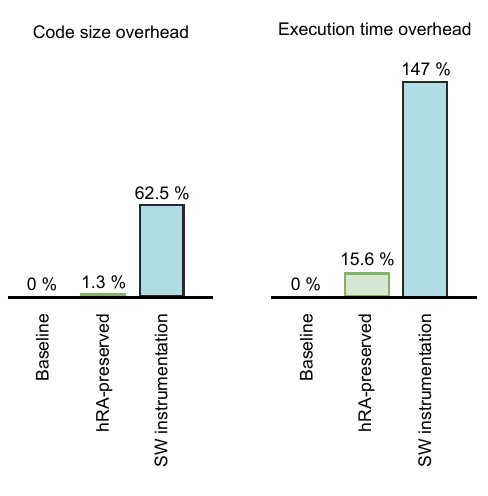}
\caption{hRA overhead compared with software instrumentation overhead. The minimal
overhead is attributable to the fact that control-flow preservation prevents the
optimizing compiler from applying some of its transformations, in order to keep the
control flow observable by means of embedded trace.}
\label{fig:hraoverhead}
\end{figure}

The residual cost is therefore of a different nature than that of instrumentation. On the
workload and target introduced in Section~\ref{sec:instr-cost} --- SQLite on a Cortex-A53
at 1.5\,GHz, measured in March 2026 --- the hRA-preserved build showed \textbf{+1.3\,\%
code size} and \textbf{+15.6\,\% execution time} against the same uninstrumented baseline
for which gcc/gcov instrumentation showed +62.5\,\% and +147\,\%
(Fig.~\ref{fig:hraoverhead}). As with the figures in Section~\ref{sec:instr-cost}, these
values are indicative for one workload and configuration rather than general benchmarks.
What is structural rather than workload-dependent is their \emph{origin}: the overhead
stems from locally preventing selected compiler transformations at the chosen decision
points, not from runtime counter traffic on the target.

\subsection{Evidence artifacts and assurance considerations}

Any modification of a build can in principle affect timing, and it would be wrong to claim
otherwise. hRA therefore aims at a \emph{release-like} measurement artifact rather than an
identical one: it avoids on-target counters and on-target coverage state, and it confines
changes to minimal, local, semantics-preserving observability scaffolding at decision and
condition boundaries. The trace acquisition path itself remains passive; the residual
on-target footprint introduced by the scaffolding is minimal, but it is not zero, and it
should be characterised for the platform in question rather than assumed away.

For audit readiness and for integration with partner toolchains, an hRA toolchain should
emit explicit, reviewable artifacts alongside the binary. Three of them carry most of the
assurance argument:

\begin{enumerate}
  \item a \textbf{transformation report} listing all modified source locations and the
        rules applied;
  \item a \textbf{build fingerprint} --- compiler version and flags --- that binds the
        mapping to one specific binary; and
  \item \textbf{trace-derived execution evidence} with stable anchors that can be
        independently reprocessed.
\end{enumerate}

Taken together these artifacts support a transparent assurance argument along three lines:
source intent is preserved, the observation mechanism does not depend on modifying runtime
behaviour to record it, and the mapping from trace to source elements is reproducible by a
third party.

\section{Hyper Coverage: Consolidation Across Test Levels and Variants}
\label{sec:hyper}

\subsection{Two consolidation problems}

Large embedded projects encounter two distinct consolidation problems, which are easily
conflated.

The first concerns \textbf{multiple test levels and harnesses}. Unit tests may run in a
dedicated environment while system tests run under customer-specific harnesses. The
resulting coverage evidence exists in different formats, produced by different toolchains,
with no common view.

The second concerns \textbf{code variants and configurations}. Even within system testing,
multiple variants arise from preprocessor and build-option differences. Classical coverage
summaries are reported per variant, and a per-function view that appears green for each
variant individually does not imply that the underlying source file is complete across the
set of variants that actually ship.

\subsection{A unifying view}

The Hyper Coverage function in TESSY\textregistered{} addresses both problems as a single
reporting and analysis concept. It consolidates evidence across test levels, combining
unit and component evidence with integration and system evidence; and it consolidates
across runs, configurations and variants, thereby revealing source-file lines that remain
untested in any relevant variant.

\begin{figure}[htbp]
\centering
\includegraphics[max width=\textwidth]{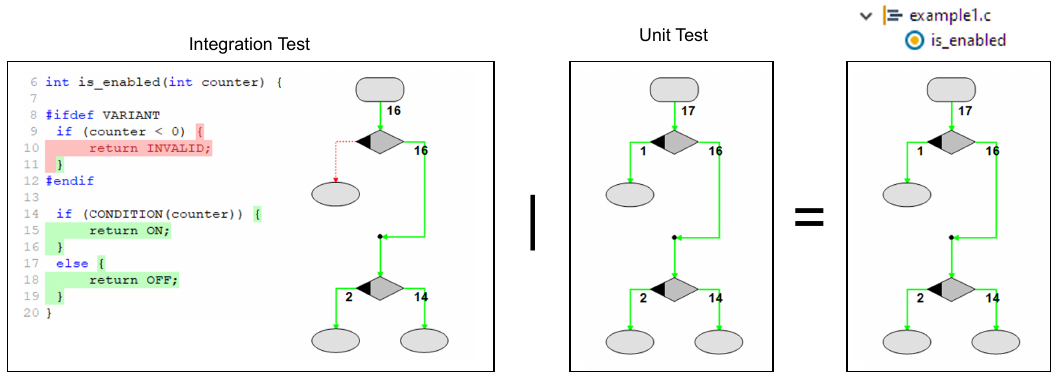}
\caption{The Hyper Coverage function in TESSY unifies evidence from unit and integration
tests.}
\label{fig:hyperunified}
\end{figure}

In practice, integration and system coverage can be exchanged through common formats for
line and branch evidence --- LCOV, for example --- and complemented by structured evidence
artifacts for decisions, conditions and MC/DC where these are required. Unit coverage is
typically available natively within the unit-test environment.

\begin{figure}[htbp]
\centering
\includegraphics[width=0.42\textwidth]{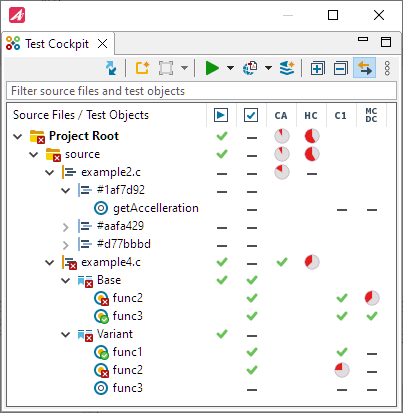}
\caption{TESSY's Hyper Coverage test cockpit: the merged results of integration and unit
tests are shown clearly.}
\label{fig:cockpit}
\end{figure}

\subsection{Consolidation at source-line level}

In variant-rich product lines not every compile-time configuration is equally relevant for
a given product and assurance boundary. We therefore begin by defining a project-specific
set $V$ of relevant variants and configurations: those builds that are in scope for
shipment, deployment, or safety and security assessment. Hyper Coverage is computed over
$V$, and any exclusion from $V$ must itself be documented.

Coverage is then consolidated at the line level of the original source file, and each line
receives one of three verdicts:

\begin{itemize}
  \item \textbf{Passed} --- all relevant coverage results for the line, across variants,
        satisfy the selected criteria.
  \item \textbf{Failed, partial} --- the line has coverage in some cases, but at least one
        required coverage measure remains below 100\,\% in at least one relevant variant
        or run.
  \item \textbf{Failed, none} --- no coverage was observed for the line at all.
\end{itemize}

The value of this formulation is that it exposes untested source lines even where the
per-variant, per-function views each appear fully covered.

\subsection{Reporting and audit readiness}

For an audit, three properties dominate: repeatability, traceability, and clear
justification records for every exclusion. Consolidated reports should consequently be
available both in human-readable form for review and in machine-readable form for pipeline
integration.

The overall process design --- the separation of test levels, the systematic handling of
coverage gaps, and the insistence on traceability for exclusions --- is aligned with the
intent of established functional safety standards: DO-178C~\cite{do178c} for airborne
software, ISO~26262~\cite{iso26262} for automotive systems, and IEC~61508~\cite{iec61508}
for generic E/E/PE safety-related systems. The integration-first coverage strategy and the
Hyper Coverage consolidation described here are intended as a practical, tool-supported
realisation of the structural coverage analysis and justification principles that those
standards embody.

\section{Deciding When Unit Tests Are Still Needed}

The integration-first approach reduces the need for audit-grade unit tests; it does not
remove unit testing. Unit tests remain necessary when

\begin{itemize}
  \item a requirement cannot be stimulated safely or economically on the integrated
        system;
  \item a physical constraint prevents exploration of a state space in integration tests,
        as with sensor ranges or mechanical limits;
  \item a module is reused across several products and requires independent evidence
        beyond a single integration context; or
  \item gap analysis identifies a decision or condition that cannot be exercised at
        integration level without unacceptable risk.
\end{itemize}

In each of these cases the unit test is added \emph{with a documented justification} and
is linked to the specific gap it is intended to close. This linkage is what prevents the
common failure mode in which unit tests silently substitute for --- and thereby mask ---
missing integration tests.

\section{Limitations and Practical Prerequisites}

The approach assumes that embedded trace is available and accessible during integration
testing. In practice this is a programme-level decision taken early: trace pins must be
provided and routed, debug access policies must permit their use, and it must be possible
to stream trace off the device without perturbing system behaviour. Retrofitting these
properties late in a programme is expensive and sometimes impossible.

Reliable object-to-source mapping further requires a stable and reproducible build
environment. Compiler versions, optimization settings and binary generation parameters
must be controlled so that the transformation from source to object code remains
deterministic and analysable. Finally, as with any measurement technology used in
regulated projects, project-specific assurance cases and tool qualification considerations
apply, and their extent depends on the intended use and the certification context.

\section{Conclusion and Outlook}

Coverage is most valuable when it reflects what the shipped system actually executes. For
embedded systems this argues for deriving evidence from integration and system tests on
representative builds --- yet instrumentation-based measurement frequently breaks
precisely the realism and the resource budgets that make those tests worth running.

This paper has presented an integration-first process together with a trace-based
technical realisation that makes code coverage measurable on integration tests running on
the device under test. We discussed why trace-only mapping fails for decision and
condition evidence under optimization, and introduced hybrid runtime analysis as a
practical mechanism that preserves observability and renders the mapping verifiable, which
in turn makes robust MC/DC evidence attainable on optimized binaries. Finally we described
Hyper Coverage as a consolidation layer spanning test levels, variants and configurations,
which reveals untested source lines and produces reports suited to audit.

The practical next step is deeper toolchain integration. We position hRA as a build-time
observability layer that complements existing coverage and test environments by combining
representative execution with verifiable decision and condition evidence on optimized
binaries, and that is designed to fit into established industrial toolchains while
preserving their existing reporting workflows.

\section*{Acknowledgements}

TRISTAN received funding from the Key Digital Technologies Joint Undertaking (KDT~JU)
under project ID 101095947 and from the German Federal Ministry of Education and Research
(BMBF), project ID 16MEE0273.

We thank Martin Heininger (Heicon Ulm GmbH) for his valuable advice and suggestions.

\vspace{6pt}
\noindent\footnotesize
Arm and CoreSight are trademarks of Arm Limited (or its subsidiaries) in the US and/or
elsewhere. AURIX and TriCore are trademarks of Infineon Technologies AG. TESSY is a
registered trademark of Razorcat Development GmbH. CEDARtools is a trademark of Accemic
Technologies GmbH. All other product names, company names and trademarks mentioned in this
paper are the property of their respective owners.

\end{document}

%% file: listings-mcdc.tex
\definecolor{hlanchorbg}{RGB}{255,247,179}

\newcommand{\hlanchor}[1]{{\setlength{\fboxsep}{1pt}%
  \colorbox{hlanchorbg}{\ttfamily\footnotesize #1}}}

\definecolor{annotgreenfg}{RGB}{23,124,49}
\definecolor{annotredfg}{RGB}{178,24,24}
\newcommand{\annotgreen}[1]{{\normalfont\footnotesize\bfseries\color{annotgreenfg}$\leftarrow$ #1}}
\newcommand{\annotred}[1]{{\normalfont\footnotesize\bfseries\color{annotredfg}$\leftarrow$ #1}}

\lstdefinestyle{mcdcasm}{
  basicstyle=\ttfamily\footnotesize,
  columns=fullflexible,
  keepspaces=true,
  showstringspaces=false,
  frame=none,
  aboveskip=2pt,
  belowskip=2pt,
  escapeinside={(*}{*)},
  literate={->}{{$\rightarrow$}}2,
}

%% file: main.bbl
\begin{thebibliography}{99}

\bibitem{dx2024}
A.~Weiss, A.~Schulz, M.~Heininger, M.~Sachenbacher, and M.~Leucker,
``Achieving Complete Structural Test Coverage in Embedded Systems Using Trace-Based
Monitoring,'' in \emph{35th International Conference on Principles of Diagnosis and
Resilient Systems (DX 2024)}, I.~Pill, A.~Natan, and F.~Wotawa, Eds., Open Access Series
in Informatics (OASIcs), vol.~125. Dagstuhl, Germany: Schloss Dagstuhl --
Leibniz-Zentrum f\"ur Informatik, 2024, p.~19:1--19:12.
doi:~10.4230/OASIcs.DX.2024.19.

\bibitem{chilenski2001}
J.~J. Chilenski, ``An investigation of three forms of the modified condition decision
coverage (MC/DC) criterion,'' Office of Aviation Research, 2001.

\bibitem{chilenski1994}
J.~J. Chilenski and S.~P. Miller, \emph{Applicability of Modified Condition/Decision
Coverage to Software Testing}, 1994.

\bibitem{lauterbach2023}
Lauterbach GmbH, ``New Multi-Mode MC/DC Coverage,'' 2023. [Online]. Available:
\url{https://support.lauterbach.com/news/posts/new-multi-mode-mcdc-coverage}

\bibitem{gcov}
\emph{gcov --- a test coverage program}, GNU Project. [Online]. Available:
\url{https://gcc.gnu.org/onlinedocs/gcc/Gcov.html}

\bibitem{trace2022}
T.~B. Preu\ss{}er, S.~Gautham, A.~D. Rajagopala, C.~R. Elks, and A.~Weiss,
``Everything You Always Wanted to Know About Embedded Trace,'' \emph{Computer}, vol.~55,
no.~2, pp.~34--43, Feb.~2022, doi:~10.1109/MC.2021.3098965.

\bibitem{coresight}
\emph{CoreSight Architecture Specification}, Arm Ltd., 2022. [Online]. Available:
\url{https://developer.arm.com/documentation/ihi0029/latest}

\bibitem{cedartools}
Accemic Technologies GmbH, ``CEDARtools.Coverage: Trace-Based Live Coverage Measurement
for Embedded Systems,'' 2025. [Online]. Available: \url{https://accemic.com/cedartools/}

\bibitem{dwarf5}
DWARF Debugging Information Format Committee, ``DWARF Debugging Information Format
Version~5,'' 2017. [Online]. Available: \url{https://dwarfstd.org/doc/DWARF5.pdf}

\bibitem{rapicover}
Rapita Systems Ltd., ``RapiCover -- Low-overhead coverage analysis for critical
software,'' 2026. [Online]. Available:
\url{https://www.rapitasystems.com/products/rapicover}

\bibitem{vector2022}
Vector Informatik GmbH and iSYSTEM AG, ``Source Code and/or Object Code Coverage for
Evaluating Safety,'' 2022. [Online]. Available:
\url{https://www.vector.com/int/en/events/global-de-en/webinar-recordings/2022/source-code-and-or-object-code-coverage-for-evaluating-}

\bibitem{do178c}
\emph{DO-178C, Software Considerations in Airborne Systems and Equipment Certification},
RTCA, Dec.~13, 2011.

\bibitem{iso26262}
\emph{ISO 26262:2018, Road vehicles -- Functional safety}, International Organization for
Standardization, 2018.

\bibitem{iec61508}
\emph{IEC 61508:2010, Functional safety of electrical/electronic/programmable electronic
safety-related systems}, International Electrotechnical Commission, 2010.

\end{thebibliography}
